\documentclass[modern]{aastex61}

\newcommand{\jgra}{{\jgr}}

\received{15 November 2017} \revised{20 December 2017}
\accepted{22 December 2017}
\submitjournal{ApJL}

\shorttitle{Self-generation of Small-scale Flux Ropes}
\shortauthors{Zheng and Hu} 

\begin{document}

\title{Observational evidence for self-generation of small-scale magnetic flux
ropes from intermittent solar wind turbulence}

\correspondingauthor{Qiang Hu} \email{qiang.hu.th@dartmouth.edu,
qh0001@uah.edu}

\author{Jinlei Zheng}
\affil{Department of Space Science \\
The University of Alabama in Huntsville \\
Huntsville, AL 35805, USA}

\author[0000-0002-7570-2301]{Qiang Hu}
\affiliation{Department of Space Science and CSPAR \\
The University of Alabama in Huntsville \\
Huntsville, AL 35805, USA}




\begin{abstract}
We present unique {and additional} observational evidence for the
self-generation of small-scale coherent magnetic flux rope
structures in the solar wind. Such structures with durations
between 9 and 361 minutes are identified from Wind in-situ
spacecraft measurements through the Grad-Shafranov (GS)
reconstruction approach. The event occurrence counts are on the
order of 3,500 per year on average and have a clear solar cycle
dependence. We build a database of small-scale magnetic flux ropes
from twenty-year worth of Wind spacecraft data. We show a
power-law distribution of the wall-to-wall time corresponding well
to the inertial range turbulence, which agrees with
 relevant observation and numerical simulation results. We
also provide the axial current density distribution from the
GS-based observational analysis which yields non-Gaussian
probability density function consistent with numerical simulation
result.
\end{abstract}




\section{Introduction} \label{sec:intro}
Small-scale magnetic flux ropes in the solar wind of durations
ranging from a few minutes to a few hours have been identified
from in-situ spacecraft data and studied for decades
\citep[e.g.,][]{Moldwin1995,Moldwin2000,Cartwright2010,Feng2008,Yu2014}.
They possess some similar features in magnetic field
configurations to their large-scale counterparts, the magnetic
clouds (MCs) of durations lasting for a dozen hours up to a few
days, but differ in certain plasma properties. Unlike MCs which
have a clear solar origin related to coronal mass ejections
(CMEs), the origin of these small-scale magnetic flux ropes is
still debated. One view is that they also have a solar source
correspondence, especially for intermediate-sized flux ropes
\citep[e.g.,][]{Feng2008}, based on some similar statistical
properties as MCs. As advocated by \citet{Borovsky2008}, the
``flux tubes" bounded by discontinuities may be rooted on the Sun,
permeating the whole interplanetary space. The other view,
supported not only by observational analysis, but also extensively
by numerical simulations over a wide range of scales
\citep{2008PhRvL.100i5005S,Greco2008,Greco2009a,Greco2009b,2010P&SS...58.1895G,2013PhPl...20d2307W,2015PhRvL.114q5002W},
states that the generation process of small-scale  magnetic flux
ropes or islands in strict two-dimensional (2D) configuration is
intrinsic to magnetohydrodynamic (MHD) turbulence often
approximated by a quasi-2D model or containing a dominant 2D
component \citep[e.g.,][and references
therein]{2007ApJ...667..956M,Zank2017}. They are believed to be
the byproduct of solar wind turbulence dynamic evolution process,
resulting in the generation of coherent structures including
``{\em small random current}", ``{\em current cores}" and ``{\em
current sheets}" \citep{Greco2009a} over the inertial range length
scales.

Accompanying these studies associated with small-scale flux ropes,
the observational analysis on the discontinuities or current
sheets possibly bounding the flux ropes (or flux tubes; see, e.g.,
\citet{Borovsky2008}) is also ongoing. Various approaches have
been utilized to identify these structures mostly as incremental
changes in magnetic field from in-situ time-series data
\citep[e.g.,][]{Bruno2001,Vasquez2007,Borovsky2008,Greco2009a,2010P&SS...58.1895G,Miao2011,Osman2014,2016ApJ...823L..39G}.
However there generally lacks a synergy between the analyses of
these two types of coherent structures, and the analysis method
for the small-scale magnetic flux ropes is outdated. In this
Letter, we explore the application of the Grad-Shafranov (GS)
reconstruction technique to the automated detection of small-scale
magnetic flux ropes for the first time. Meanwhile, we report on
the successful identification of an unprecedented number of the
small-scale flux rope events via the GS approach and associated
analysis results including unique physical characterization of
these structures, {especially the axial current density
distribution}, which enables a direct comparison with the
numerical simulation results for 2D MHD turbulence. {To the best
of our knowledge, the estimate of current density has to be
 achieved by using closely-spaced multiple spacecraft  through the curlometer approach, such
 as from
the Cluster and Magnetospheric Multiscale  (MMS) missions
\citep[see, e.g.,][for such a comparison of current density
between MMS measurements and 2D
simulations]{2018SSRv..214....1G}.}

The GS reconstruction technique is based on the GS equation,
describing 2D (or 2.5D with non-vanishing axial magnetic field)
cylindrical magnetic field and plasma configurations in
magnetohydrostatic equilibrium (see \citet{Hu2017b} for a
{comprehensive} review) {that is more general than the force-free
assumption}. For the magnetic field components
$\mathbf{B}=(\partial A/\partial y, -\partial A/\partial x,
B_z(A))$ defined by the 2D magnetic flux function $A(x,y)$ in a
Cartesian coordinates (with $z$-axis along the flux rope axis, and
$\partial/\partial z\approx 0$), the transverse force balance
yields the GS equation {(reduced from the usual equation $\nabla
p=\mathbf{J}\times\mathbf{B}$)},
\begin{equation}
\frac{\partial^2 A}{\partial x^2}+\frac{\partial^2 A}{\partial
y^2}=-\mu_0\frac{dP_t}{dA}=-\mu_0 J_z(A).\label{eq:GS}
\end{equation}
Here the right-hand side gives the axial current density $J_z$
which is a total derivative of the transverse pressure
$P_t=p+B_z^2/2\mu_0$, the sum of the plasma pressure and the axial
magnetic pressure, with respect to $A$. All these quantities can
be evaluated along a single-spacecraft path across a flux rope
structure. {{Since the flux function $A$ characterizes the nested
cylindrical flux surfaces of a flux rope, the axial current
density distribution throughout such a flux rope configuration is
readily obtained by the function $J_z(A)$ determined from in-situ
spacecraft data.}} Additionally the cross section of a cylindrical
flux rope given by the solution $A(x,y)$ to the GS equation over a
rectangular domain can also be obtained numerically
\citep{Hu2001,Hu2002}.  The technique has been widely applied to
reconstruct structures in a variety of space plasma regimes
\citep[see, e.g.,][]{Hu2017b}. The application of the GS method to
the small-scale structures of relevance to the quasi-2D MHD
turbulence as envisaged by \citet{2007ApJ...667..956M} has just
begun.

{\citet{2016ApJ...826..205T} took yet another approach by
evaluating the MHD rugged invariants \citep[see also,
][]{2013ApJ...776....3T} within about 144 flux ropes identified in
prior studies. They concluded that flux ropes represent
well-organized structures coming from the dynamical evolution of
MHD turbulent cascade, in which the MHD invariants are
inter-related. They further stated that the flux ropes dynamically
evolve toward a final steady state in which the (normalized)
magnetic helicity $\sigma_m$, and cross-helicity $\sigma_c$ values
within the structures are distributed according to
$\sigma_m^2+\sigma_c^2=1$.
  We expect to
complement that study by providing a more exhaustive list of
events equipped with more comprehensive characterizations of flux
rope properties. For instance, from the GS reconstruction output,
we will be able to derive, quantitatively and directly, the total
magnetic energy, magnetic flux, and (relative) magnetic helicity
\citep{Hu2017b} contained in each flux rope, although this is not
the focus of the current analysis, and we have excluded Alfv\'enic
structures or structures with high Alfv\'enicity.}

We present and introduce briefly the small-scale flux rope
database built by an automated process based on the GS method in
the next section. In Section~\ref{sec:w2w}, we present the
analysis results of the wall-to-wall time and the axial current
density distributions derived from our database and compare
directly with \citet{Greco2009a}. We conclude in the last section,
and signify the uniqueness of our approach and result in support
of the view of the self-generation of small-scale flux ropes  via
MHD turbulence.

\section{Small-scale Flux Rope Database via the GS Reconstruction Method} \label{sec:database}
We have built a small-scale magnetic  flux rope database via the
GS reconstruction method. The GS method was first applied to
reconstruct cross sections of small-scale magnetic flux ropes in
the solar wind with durations of about half an hour in
\citet{Hu2001}. In the present study, we apply to Wind in-situ
spacecraft measurements of 1-minute cadence between 1996 and 2016,
based on largely the same principles and  procedures without
carrying out the final numerical reconstruction of solving for a
2D solution to the GS equation on the cross-sectional plane.
Detailed descriptions of the automated detection algorithm
including a flowchart illustrating the step-by-step implementation
of the GS-based algorithm are presented in \citet{zheng_diss}.

\begin{table}[htb]
    \caption{Metrics and Selection Criteria for the GS-based Automated Flux Rope Detection Algorithm}\label{tbl:metrics}
    \centering
    \begin{tabular}{ccccc}
        \hline
        Duration (minutes) & $\bar{{B}}$ (nT) & $RES$ & $R_{f}$ & Wal\'en  slope\\
        \hline
        $9\sim361$  & $\geq5$  & $\leq0.12$ & $\leq0.14$ & $\leq0.3$\\
        \hline
    \end{tabular}
\end{table}
In short, the detection procedures start by sliding a rectangular
window of a chosen width through the time series to select the
data interval for analysis. For each interval and the selected
magnetic field and plasma parameters, the same procedures apply as
the standard GS reconstruction of 2.5D magnetohydrostatic
structures. The main steps are to obtain the flux function $A$ and
the transverse pressure $P_t$ values along the spacecraft path in
a properly determined trial frame of reference \citep{Hu2002}.
Then the necessary conditions for a 2.5D cylindrical magnetic flux
rope configuration, mainly the requirement that the function
$P_t(A)$ be single-valued based on the GS equation and the
double-folding pattern in $A$ across the flux rope characteristic
of nested closed flux surfaces (each of a distinct $A$ value), are
checked by producing a set of quantitative metrics. After a
trial-and-error process, especially by enumerating all the
possible $z$-axis orientations in its parameter space, the set of
metrics is obtained and a flux rope candidate is positively
identified if the selection threshold conditions are satisfied, as
indicated in Table~\ref{tbl:metrics}. The flux rope interval is
then identified and recorded with a size/duration limited by the
width of the data window. Flux ropes with different sizes are
simply identified by iterations, repeating the aforementioned
procedures with different-width sliding windows.
Additional post-processing procedures are taken to clean up
overlapping intervals and to ensure good-quality events in our
database according to the additional metrics in
Table~\ref{tbl:metrics}. In summary, the two metrics, $RES$ and
$R_{f}$, originally defined in \citet{Hu2002} and \citet{Hu2004},
respectively, evaluate the quality of $P_t(A)$ being single-valued
and each has to be smaller than the specific threshold. The
Wal\'en slope \citep{Paschmann2008}, corresponding to the average
ratio between the remaining plasma flow (ideally zero) in the
frame of reference moving with the flux rope and the local
Alfv\'en velocity, is used as a criterion to remove Alfv\'enic
structures. We also have the option to restrict our database to
only contain flux ropes of average magnetic field magnitude
$\bar{{B}}\geq 5$ nT, following the previous studies
\citep[e.g.,][]{Cartwright2010} and avoiding complications due to
low Alfv\'en speed, in the present study. {{Therefore, our
database contains the small-scale flux ropes with helical magnetic
field configuration and negligible remaining plasma flow,
corresponding to $\sigma_c\approx 0$ and the maximum
$\sigma_m\approx\pm 1$ as expected based on
\citet{2016ApJ...826..205T}. However our database contains events
with a wide range of plasma (proton) $\beta$ values, yielding both
a mean and a median around 0.5, since we are not limited to
finding low-$\beta$ structures only.}}

\begin{figure}
 \centerline{\includegraphics[width=.6\textwidth,clip=]{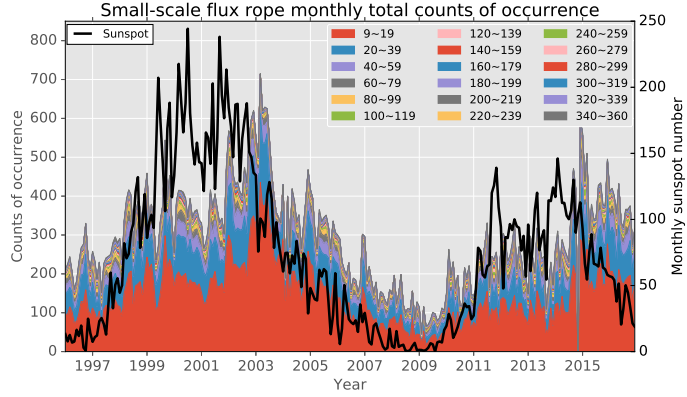}}
 \caption{Monthly counts (left axis) of small-scale magnetic flux rope events identified from Wind in-situ spacecraft
 measurements during 1996-2016. The counts are separated for flux ropes of different durations and color-coded as indicated
 by the legends in minutes. The corresponding monthly sunspot numbers are given by the black curve (right axis).   }\label{fig:counts}
 \end{figure}
We have identified a total number of 74,241 distinct small-scale
magnetic flux rope events with durations ranging from 9 to 361
minutes from the Wind spacecraft data sets. We have summarized and
compiled our detection results onto a publicly available website,
\url{http://fluxrope.info}, which contains some essential
information about the properties of the detected flux rope
structures. Various statistical analyses have been performed and
reported in \citet{zheng_diss}. At a glimpse,
Figure~\ref{fig:counts} presents the monthly event counts in our
database, covering the past two solar cycles. Clearly the total
counts including all events of variable durations follow the
monthly sunspot numbers, hinting at solar-cycle dependency of
their occurrence. The events of smaller durations generally have
greater rates of occurrence.
 The peaks of occurrence
counts tend to appear in the declining phase of each solar cycle.
On average, we have identified more than 3,500 small-scale
magnetic flux ropes per year. {We caution, however, against a
quick conclusion about the solar origin of these structures,
because we believe there should be a distinction between ``solar
origin" and ``solar activity dependence" as indicated by the
dependence on sunspot numbers for the latter. In addition, some
local turbulence processes could also be modulated by the solar
activity, leading to the solar-cycle dependence as exhibited here.
For example, perhaps the simplest case was the modulation of
cosmic ray transport coefficients by the solar-cycle variations in
the magnetic field magnitude and variance \citep{MANUEL20111529}.
Furthermore, the recent works of \citet{2017ApJ...849...88Z,2018Z}
showed the solar-cycle dependence of various turbulence quantities
affecting cosmic ray diffusion, such as Els\"asser variables,
 correlation lengths, and residue energy etc. They are all
derived locally and with contributions from local driving sources,
based on sophisticated MHD turbulence theory
\citep{Zank2017,2014ApJ...793...52A}. }

\section{Wall-to-Wall Time and Axial Current Density
Distributions}\label{sec:w2w}

With such a small-scale magnetic flux rope database, we are
positioned to perform detailed statistical analysis in addition to
the occurrence rate distribution demonstrated in
Figure~\ref{fig:counts}.
We
report here  important findings about the wall-to-wall time and
the axial current density $J_z$ distributions from the identified
flux ropes. They have direct and significant relevance to the
findings of \citet{Greco2009a}.

\begin{figure}
 \centerline{\includegraphics[width=.8\textwidth,clip=]{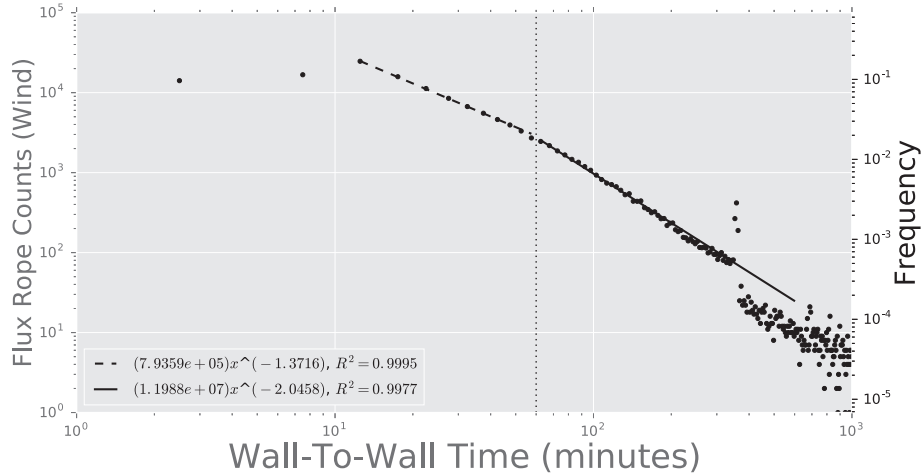}}
 \caption{The wall-to-wall time distribution of small-scale flux ropes in our database. The dashed and solid black lines
 are the power-law fittings to the two sections divided by the dotted line at 60 minutes. The fitting parameters are
 denoted,
 for each line, in the lower left corner with the goodness-of-fit parameter $R^2$, the  coefficient of determination.}\label{fig:w2w}
 \end{figure}
\begin{figure}
 \centerline{\includegraphics[width=.6\textwidth,clip=]{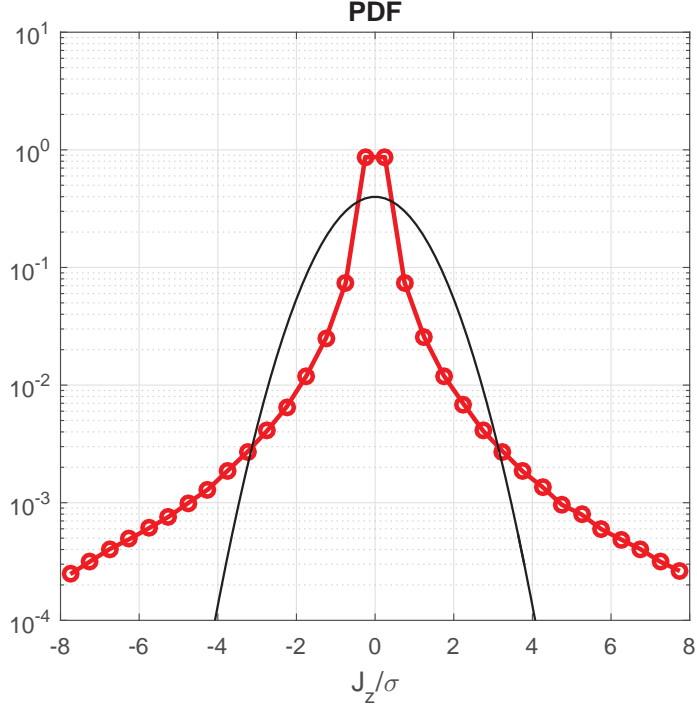}}
 \caption{The probability density function (PDF) of the axial current density $J_z$ distribution associated with magnetic flux ropes
  normalized by the standard deviation $\sigma$ ($\sigma=8.42\times 10^{-11}$ A/$\mathrm{m^2}$).
  The thin black curve is the
 standard Gaussian distribution with unit variance.}\label{fig:Jz}
 \end{figure}

The results from our analysis are presented in
Figures~\ref{fig:w2w} and \ref{fig:Jz}. The current density is
derived from $dP_t/dA$, as indicated in equation~(\ref{eq:GS}),
where both $P_t$ and $A$ are evaluated from in-situ spacecraft
measurements for each individual event. To facilitate a direct
comparison with the numerical simulation result which is 2D in
nature \citep{Greco2009a}, we calculated the axial current density
samples in each event  at a rate proportional to its scale size
and congregated the results together from all events. The flux
rope  wall-to-wall time is simply the distribution by putting
together the flux rope duration and the separation time in-between
adjacent flux rope intervals. The flux rope wall or boundary in
our database is considered as a type of discontinuity \citep[see,
e.g.,][]{Borovsky2008} or current sheet. Therefore the
wall-to-wall time in our analysis is used as a proxy to the
waiting time between current sheets or discontinuities of
negligible thicknesses. Such waiting time distributions (WTDs)
have been analyzed in a number of previous works
\citep[e.g.,][]{Bruno2001,Vasquez2007,Greco2008,Greco2009a,Greco2009b,Miao2011}
by directly identifying current sheets or discontinuities from
in-situ time-series data. We note that our proxy is unique and
valid for flux ropes bounded by discontinuities broadly defined as
locations where the magnetic field and/or plasma parameters
change. This is consistent with our choice of flux rope boundaries
based on the GS method \citep{Hu2004}. {They are chosen as the
locations along the approximately single-valued and double-folded
$P_t$ versus $A$ curve, corresponding to the specific data points
in the time series of both the magnetic field and plasma
measurements. Beyond those points, i.e., beyond the flux rope
boundary, the data, in terms of $P_t=p+B_z^2/2\mu_0$ as a function
of $A$, start to deviate. This often corresponds to the concurrent
changes or increments in magnetic field and plasma parameters when
a flux-tube (or rope) boundary is encountered
\citep{Borovsky2008}.} Therefore this lends to the justification
for using the wall-to-wall time as a proxy to the current sheet
waiting time.

Figure~\ref{fig:w2w} shows the wall-to-wall time distribution of
small-scale flux ropes. The distribution is well fitted by two
power law functions of different power indices as indicated, with
a break point located at $\sim60$ minutes. The outliers near the
two ends are  due to  cutoff effect of limited durations for these
events. The WTDs of the observed solar wind discontinuities in
\citet{Greco2009a} showed a power law within the break point at
$\sim$50 minutes, corresponding to the typical correlation length
scale \citep{Matthaeus2005}, which is consistent with our result.
In addition, the power law index from \citet{Greco2009b} is $-0.92
\pm 0.03$ from numerical simulation, and $-1.23 \pm 0.03$ from the
in-situ spacecraft observations at 1 AU, for the range below the
break point, i.e., in the inertial range of solar wind turbulence.
They compare well with our result, $-1.37$ for the dashed black
line in Figure~\ref{fig:w2w}. We note that for the case including
additional flux rope events by relaxing the criterion on $\bar{B}$
to $\bar{B}>0$, the number of events doubles and the power law
fitting to the wall-to-wall time persists. The index for the
inertial range becomes $-1.22$. The section beyond the break point
can still be fitted by a power law function with a power index
$\sim -2$. However the interpretation of this portion is not
certain and has yet to be improved with better statistics.

 Figure~\ref{fig:Jz} shows the PDF of the normalized axial current
 density distribution inside and near the boundaries of the
 small-scale magnetic flux ropes in our database, in the same
 format as the result presented by \citet{Greco2009a} from 2D MHD
 simulation. They also characterized their non-Gaussian PDF by three
 regions: regions I and III are super-Gaussian, around the center and toward the tails of the PDF, corresponding to
 current-sheet like structures in-between magnetic flux ropes or
 magnetic islands, but with the weakest and strongest current
 density in magnitude, respectively; region II corresponds
 to the sub-Gaussian  section with modest magnitude of current
 density, mostly appearing inside flux rope cores. Our result in
 Figure~\ref{fig:Jz} exhibits the same general non-Gaussian
 features in agreement with \citet{Greco2009a}.

\section{Conclusions}\label{sec:con}
In conclusion, in this Letter, we report the analysis result on
the wall-to-wall time and axial current density distributions from
in-situ spacecraft observations of small-scale magnetic flux ropes
based on an extensive event database and the unique GS model
output. Our results show a power-law distribution of the
wall-to-wall time for length scales smaller than the correlation
length, corresponding well to the inertial range of solar wind
turbulence. This result is consistent with the analysis result of
\citet{Greco2009a} from both observational analysis using an
entirely different approach and  MHD simulations of intermittent
turbulence. In addition, we also obtain the non-Gaussian
distribution of the axial current density associated with the
magnetic flux ropes, which is also in agreement with the numerical
simulation result of \citet{Greco2009a}. We therefore conclude
that we have provided {unique {and additional}} observational
evidence in support of the view of self-generation of  coherent
structures, such as small-scale magnetic flux ropes and current
sheets, locally from MHD turbulence. {In light of the analysis by
\citet{2016ApJ...826..205T}, we plan to extend our analysis to
include structures of significant remaining flow, i.e., structures
with $\sigma_c\ne 0$. This is feasible since the extension of the
GS method to the GS-type with significant field-aligned flow and
even to the 2D MHD equilibrium has been developed
\citep[see][]{Hu2017b}. Meanwhile we also invite other researchers
to extend their relevant studies by utilizing our extensive
database. }

\acknowledgments We are grateful to our colleagues, Drs.~Laxman
Adhikari, Jakobus le~Roux, Gang Li, Gary Webb, Gary Zank, and
Lingling Zhao for illuminating discussions and suggestions that
have made this work possible. We acknowledge NASA grants
NNX12AH50G, NNX14AF41G, NNX17AB85G, subawards NRL N00173-14-1-G006
and
 SAO SV4-84017, and NSF grant AGS-1650854 for support.
We thank Drs.~M.~G.~Linton and P.~Riley for stimulating
discussions on the GS-based flux rope identification which
initiated this work. {We also thank the reviewer for his/her
expert comments and suggestions.}

 \bibliographystyle{aasjournal}
 \bibliography{bibw2w}
\end{document}